\documentclass[10pt,sigconf,nonacm]{acmart}
\pdfoutput=1

\usepackage[english]{babel}
\usepackage{blindtext}

\usepackage{import}
\usepackage{graphicx}
\graphicspath{ {figs/} }
\usepackage{listings}
\usepackage{caption}
\usepackage{subcaption}
\usepackage{tabularx}
\usepackage{cprotect}

\renewcommand\footnotetextcopyrightpermission[1]{} 
\setcopyright{none}

\acmDOI{ }

\acmISBN{ }

\acmConference[ ]{}
\acmYear{}
\copyrightyear{}

\acmPrice{}

\begin{document}

\title{Examining Interplay of Compression and Encryption and Applicability to 5G Teleoperations}

\author{Duncan Joly}
\email{joly0012@umn.edu}
\affiliation{
  \institution{University of Minnesota}
  \city{Minneapolis}
  \state{MN}
  \orcid{0000-0003-3058-0758}
}

\author{Jason Carpenter}
\email{CARPE415@umn.edu}
\affiliation{
  \institution{University of Minnesota}
  \city{Minneapolis}
  \state{MN}
}

\author{Zhi-Li Zhang}
\email{zhzhang@cs.umn.edu}
\affiliation{
  \institution{University of Minnesota}
  \city{Minneapolis}
  \state{MN}
  }

\date{University of Minnesota\\Minneapolis, MN}

\begin{abstract}
Modern IoT and networked systems rely on fast and secure delivery of time-critical information. Use cases such as teleoperations require fast data delivery over mobile networks, which despite improvements in 5G are still quite constrained. Algorithms for encryption and compression provide security and data size efficiency, but come with time and data size trade-offs. The impact of these trade-offs is related to the order in which these operations are applied, and as such necessitates a robust exploration from a performance perspective. In this paper, we assess several compression and encryption algorithms, combinations of their execution order, timings and size changes from such order, and the implications of such changes on 5G teleoperations. From our assessments we have three major takeaways: (1) Compression-First is faster and more compressed, except for certain circumstances. (2) In these specific circumstances, the compression against a raw file leads to a lengthier time than if applied to an encrypted file first. (3) Applying both encryption and compression on data samples larger than 10MB is impractical for real time transmission due to the incurred delay.
\end{abstract}

\maketitle


\section{Introduction}
Large scale networking and device connectivity support the modern world. From phones to pacemakers, devices must communicate critical information over networks quickly, safely, and privately. With the increasing integration of digital components into the economy comes a rise in the amount of cybercrime committed. Malicious actors aim to take control of devices, eavesdrop on data, and in some cases, cause physical harm. With many new devices coming online\cite{Alam_2020}, this concern is quite substantial. Further, these devices often use mobile networks, such as is the case with many IoT devices and teleoperation suites. These networks often have limited resources that must be allocated between the connected devices, leading to each device receiving restrained service quality. 

Protecting the data of these devices while also staying within the restrictions of constrained networks involves the use of compression and encryption algorithms. Generally speaking, encryption algorithms secure data with the possibility of increasing its size, and compression algorithms reduce the size of data. These algorithms are often paired together to achieve some security and compression capacity. These algorithms take time to operate, and thus may complicate real-time use cases such as teleoperations. Further, the timing and size impacts of these algorithms change based on the order in which they are applied to a given piece of digital information. The current understanding is that unless there are specific attacks to guard against such as CRIME\cite{attack-definition-CRIME}, one should compress first (CF) rather than encrypt first (EF) to achieve the best compression without compromising security\cite{encrypt-compress-order-0,encrypt-compress-algorithm-1,encrypt-compress-order-3,encrypt-compress-order-4,encrypt-compress-order-5}. To further resolve this, newer algorithms aim to combine compression and encryption to achieve more harmonious results\cite{encrypt-compress-algorithm-0,encrypt-compress-algorithm-1,encrypt-compress-algorithm-2}. While there is substantial support for CF, there is not a robust exploration of the specific performance impacts of the relative order.

\subsection{Contributions}
In this paper, we assess the performance impacts of the order of application for various common compression and encryption algorithms with respect to time and data size changes. We then apply these findings to a time- and size-sensitive use case: 5G teleoperations. Broadly our findings can be summarized as follows:

\noindent{\textbf{$\bullet$ }} We confirm that Compression-First (CF) is  overall more performant than Encryption-First (EF) with the distribution of tests showing a broadly faster operation time and a compression ratio improvement generally between 25-50\%.

\noindent{\textbf{$\bullet$ }} However, we find that for file sizes larger than 10MB and some certain algorithms (\verb|bzip| and \verb|Fernet|) the faster-performing combination is actually Encryption-First with a 40-50\% time improvement.

\noindent{\textbf{$\bullet$ }} Encryption-First, under certain algorithm combinations, can out-perform Compression-First, such as with specific algorithms like \verb|bzip|, \verb|Fernet|, \verb|NaCL|, and \verb|gzip| regardless of file size.

\noindent{\textbf{$\bullet$ }} When considering a teleoperations use case, we find that 73\% of the operation pairs applied to file sizes smaller than 1MB had a total operation time of less than 100ms and thus suitable for real-time operation. Further, we find that for file sizes larger than 10MB the the operation time always exceeded the 100ms delay and thus unsuitable for real-time operation.

The code and data from this project are available at:\\
\\
\href{https://github.com/duncanjoly13/encoding-compression-investigation}{https://github.com/duncanjoly13/encoding-compres\\sion-investigation}.

\section{Background And Related Work}
In this section we will cover some background concepts and related work for understanding the interactions for compression and encryption. The essential metrics this paper will cover are data size (expressed as file sizes), compression ratio (relative size of data after operations compared to original), and operation time (time taken to compress, encrypt, decrypt, and decompress). However, a core security metric, entropy, is important for measuring the "disorder or randomness in a closed system"\cite{definition-entropy-NIST}. In other words, entropy represents how encrypted data and/or how compressible it is\cite{encrypt-compress-order-1}. Broadly, compression uses patterns for its purposes, and encryption attempts to break patterns\cite{encrypt-compress-order-1}. We note entropy as an important aspect of compression and encryption, but we save an exploration of entropy and operation order for future work. 

Broadly, there are two orders to apply compression and encryption: Compression-First (CF) and Encryption-First (EF). CF is the most common approach and is understood to be high performing while also retaining high entropy\cite{encrypt-compress-algorithm-0, encrypt-compress-order-2,encrypt-compress-order-3,encrypt-compress-order-4}. EF is not common, but may be employed to avoid certain attacks that take advantage of the size of encrypted messages such as CRIME\cite{attack-definition-CRIME, encrypt-compress-order-1, encrypt-compress-order-2}. Generally speaking, encrypted data that is easily compressed is not well encrypted\cite{encrypt-compress-order-0}, and alongside the other issues it may present, is also vulnerable to differential cryptanalysis\cite{encrypt-compress-algorithm-2}. Due to the pressure of requiring security while maintaining reasonable data sizes, newer algorithms have emerged to integrate the two operations into a cohesive algorithm\cite{encrypt-compress-algorithm-0,encrypt-compress-algorithm-1,encrypt-compress-algorithm-2}. Our work focuses on a set of common individual algorithms and saves the cutting edge algorithms for future work.

\section{Problem Formulation}

\begin{figure*}
    \centering
    \includegraphics[width=1\linewidth]{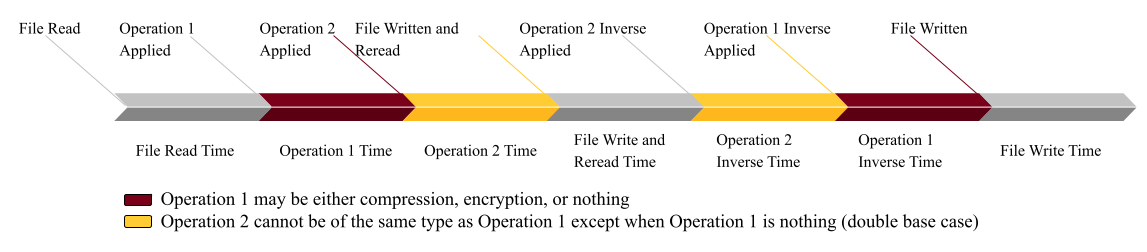}
    \caption{Compression and Encryption Pipeline.}
    \label{fig:pipeline}
\end{figure*}

To formally evaluate the timing and file size impacts of a given file’s compression and encryption life cycle, we outline a process timeline for operations (encryption or compression) as illustrated in Figure \ref{fig:pipeline}. From this process, we highlight three high level metrics: (1) Operation and Operation inverse 1\&2 times, which measures the time for a discrete compress/decompress and encrypt/decrypt. (2) Total time, which is the sum total of all operations and inverses for a particular algorithm grouping. Finally, (3) intermediate and final file sizes which provide us insights into the operation-specific impacts and overall impact on the targeted data file.

\begin{table}[]
    \centering
    \begin{tabular}{|c|c|}
        \hline
         Algorithm Name & Type \\
         \hline
         bzip2 (bz2, bzip)\cite{bzip2} & compress\\ \hline
         gzip\cite{gzip} & compress\\ \hline
         lzma\cite{lzma} & compress\\ \hline
         zipfile (zip)\cite{zipfile} & compress\\ \hline
         AES\cite{AES} & encrypt\\ \hline
         Fernet\cite{Fernet} & encrypt \\ \hline
         NaCl (XSalsa20 \cite{NaCl, XSalsa20, XSalsa_spec} with 192bit nonce) & encrypt \\ \hline
    \end{tabular}
    \caption{Evaluated Algorithms.}
    \label{tab:algs}
\end{table}

\begin{table}[]
    \centering
    \begin{tabular}{|c|c|c|}
        \hline
         File Size (B) & File Type & Data Origin \\ \hline
         85 & CSV & Vehicle GPS Sample \\ \hline
         174 & CSV & Novatel GPS Unit \\ \hline
         362 & CSV & Novatel IMU\\ \hline
         451 & CSV & Novatel enhanced GPS\\ \hline
         520 & bytes & Ouster LiDAR Telemetry Packet \\ \hline
         564 & CSV & Novatel Odometry \\ \hline
         1206 & bytes & Ouster LiDAR Data Packet  \\ \hline
         5052 & CSV & MobileEye Lane Marker Sample \\ \hline
         1086844 & text & 1MB of enwik8 \\ \hline
         10239975 & text & 10MB of enwik8 \\ \hline
         11081517 & PDF & \href{https://freetestdata.com/document-files/pdf     /}{testing PDF}\\ \hline
         101128023 & text & 95MB of enwik8 \\ \hline
    \end{tabular}
    \caption{Evaluated files: A selection of files across several domains, data formats, and compositions.}
    \label{tab:files}
\end{table}

\section{Methodology and Testing}
We examine the effect that changing the order of compression and encryption has on the file size and operation time. Using the process outlined in Fig. \ref{fig:pipeline}, we measure the time elapsed at each step of a file being encrypted and compressed, written to disk, and subsequently decrypted and decompressed. We vary the order of compression and encryption. We select 4 common compression algorithms and 3 encryption algorithms (outlined in Table \ref{tab:algs} along with 12 different file sizes (Table \ref{tab:files}). The files are taken from a selection of vehicle sensors such as LiDAR and GPS unit samples, public large and small text samples such as enwik8\cite{enwik8}, and a \href{https://freetestdata.com/document-files/pdf/}{free testing PDF}. Tests on files larger than or equal to 10MB are repeated 32 times whereas tests on files smaller than 10MB are repeated 100 times. These repetitions, totaling 39840 tests, alleviate natural variation in operation times and provide a clean average for each combination. 

\begin{figure*}[ht]
        \centering
    \includegraphics[width=0.45\textwidth]{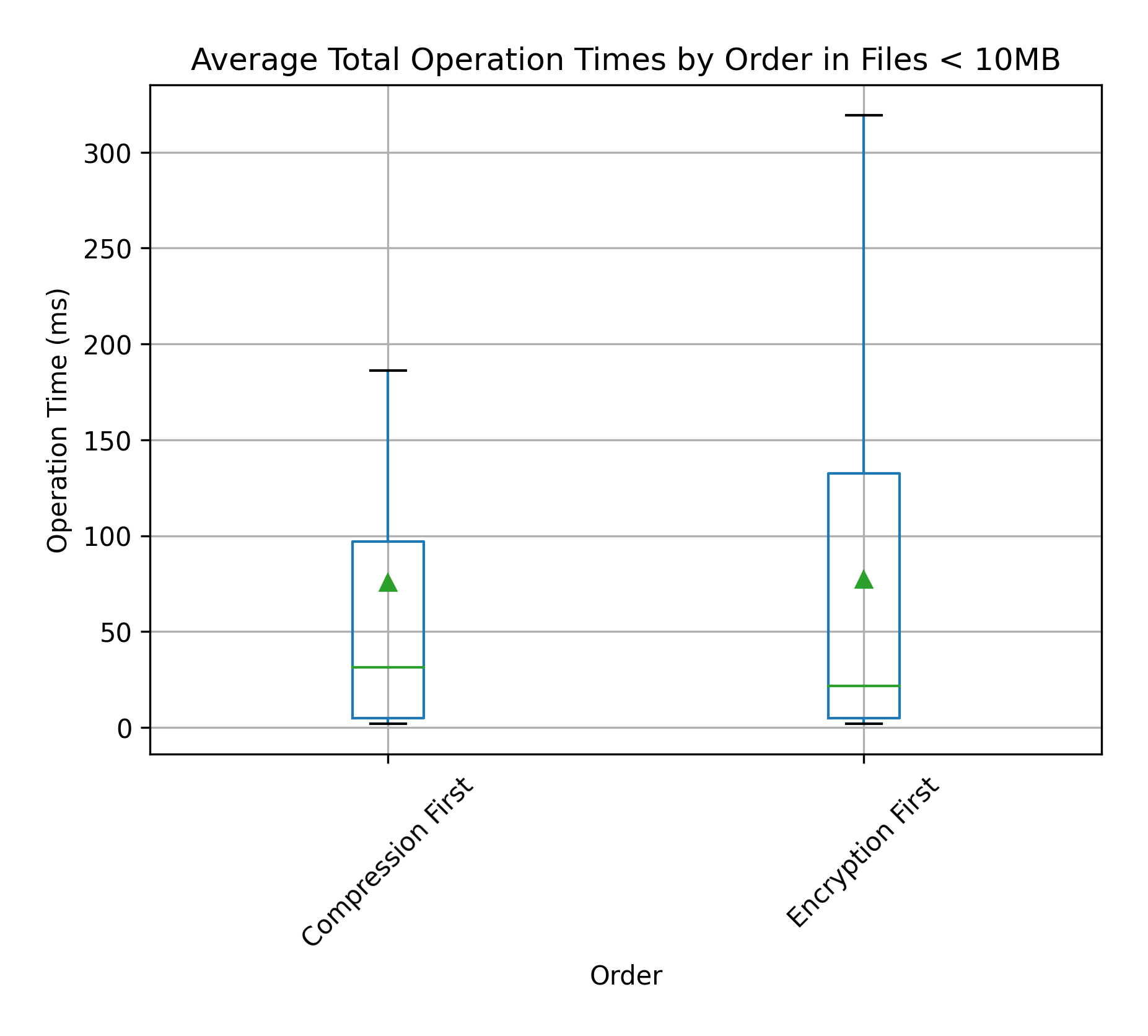}
    \includegraphics[width=0.45\textwidth]{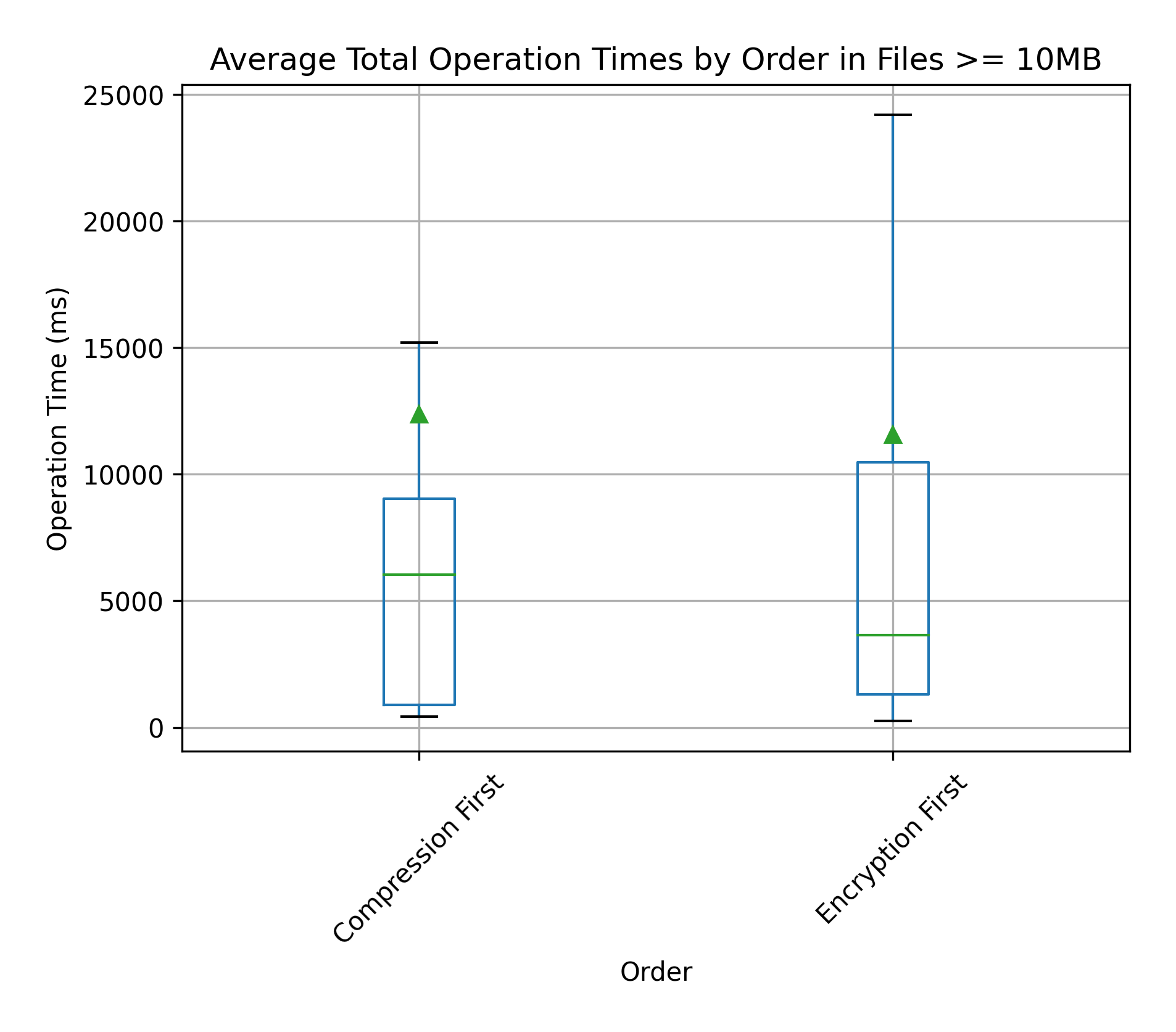}
    \caption{Operation timings: note the overall expected timings for CF and EF.}
    \label{fig:operation-timings}
\end{figure*}

\begin{figure*}[ht]
        \centering
    \includegraphics[width=0.48\textwidth]{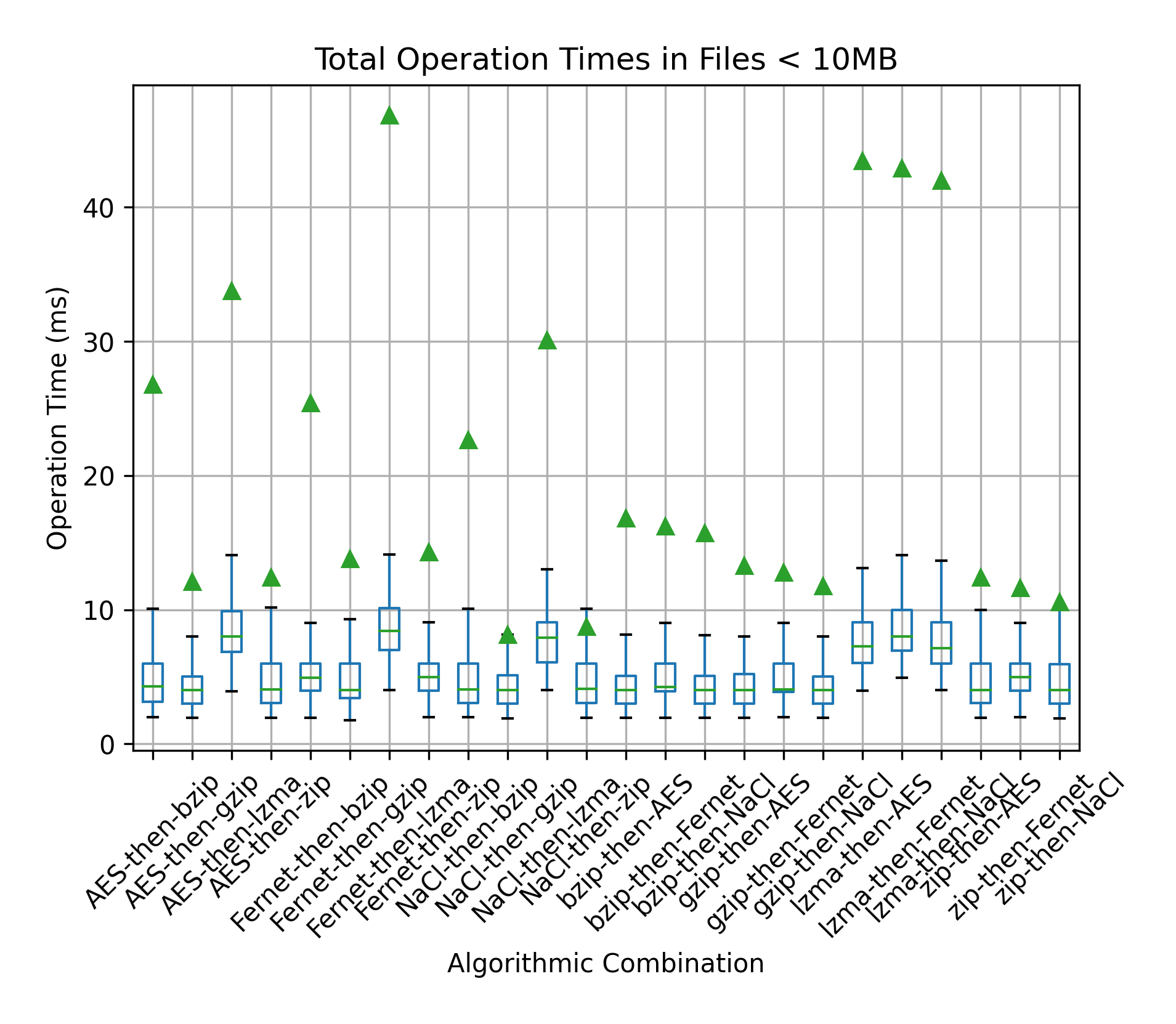}
    \includegraphics[width=0.48\textwidth]{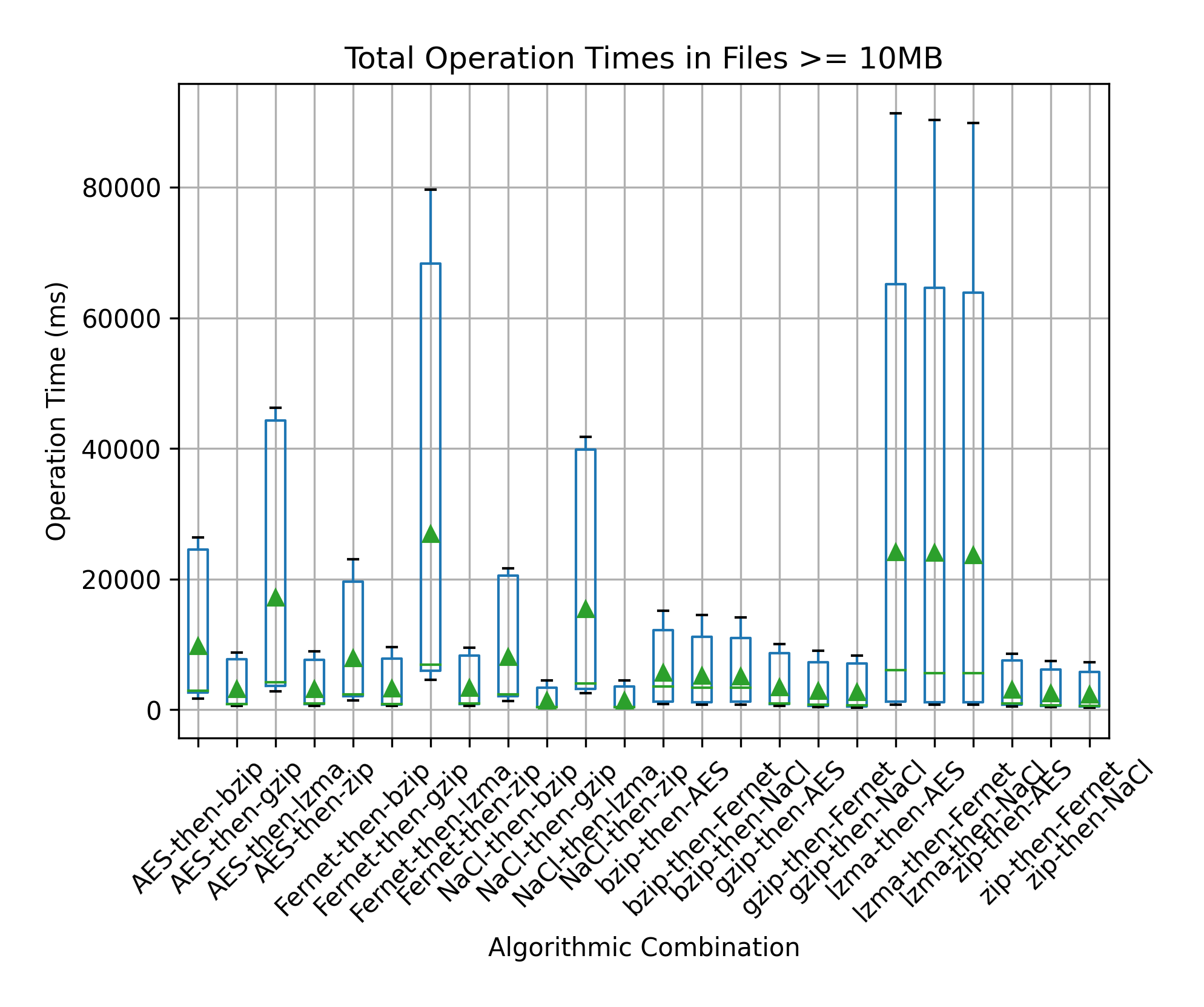}
    \caption{Operation timings by algorithm. We note the cases where EF is faster overall. We also note that the left side (EF) has competitive performance with the right side group (CF).}
    \label{fig:operation-timings-alg-specific}
\end{figure*}

\section{Results Examination}
In this section we review the results of our battery of compression and encryption operations against our test files. We first consider the timing impacts of our operations and then consider the impact order has on file size.

\subsection{Timing Impacts}
\begin{figure}
    \centering
    \includegraphics[width=0.45\textwidth]{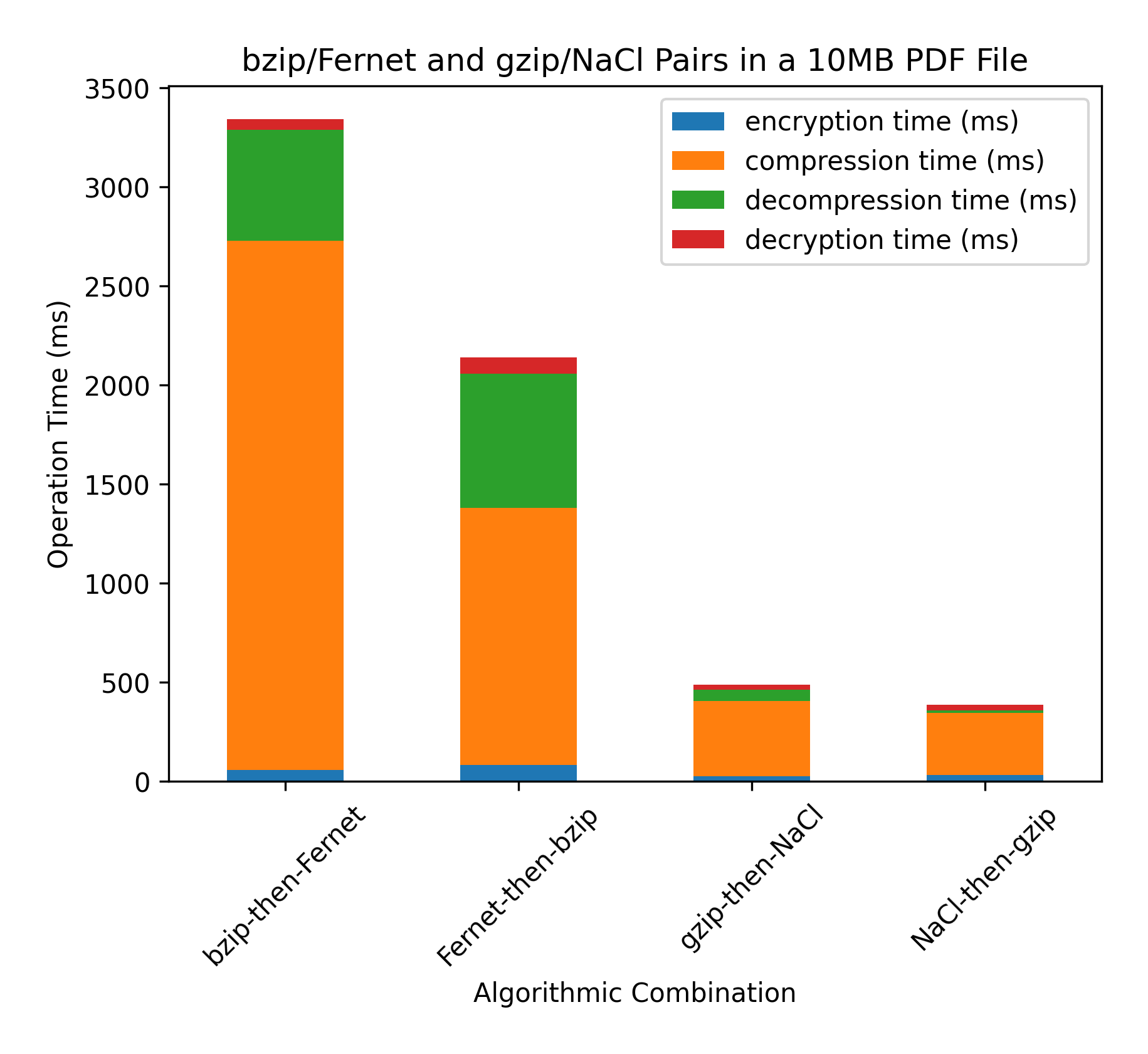}
    \cprotect\caption{Timing breakdowns of the \verb|bzip|/\verb|Fernet| and \verb|gzip|/\verb|NaCl| pairs, showing a case of Encryption-First operating faster (approx. 50\% reduction for \verb|Fernet-then-bzip| and a smaller approx. 5\% change in \verb|NaCl-then-gzip|).}
    \label{fig:barchart_bzip/fernet_gzip/nacl}
\end{figure}

Overall, we see that CF is on average slightly faster than EF; however the overall distribution of CF algorithm pairs does better than the general distribution of EF algorithm pairs. Interestingly, the same trend is somewhat complicated when considering file sizes larger than 10MB, where we see the average CF operation time take roughly 2000ms more. These results reinforce the broad consensus that CF is better in terms of performance.

Digging in deeper, we examine specific algorithm pairs split by file size in Fig. \ref{fig:operation-timings-alg-specific} with the left side algorithms as CF and the right side ones as EF. First, we see that for file sizes smaller than 10MB the overall distribution is tight around 5-6ms with the averages dominated by outliers (excluded from the figures for clarity). For file sizes larger than or equal to 10MB, we see the distributions are dominated by a handful (2-3) slower-performing algorithm pairs. If these algorithms are omitted the distributions would roughly center around 4000-5000ms.

Further, we see the specific pairs \verb|bzip|/\verb|Fernet| and\\ \verb|gzip|/\verb|NaCL| actually perform faster when encryption is applied first as opposed to Compression-First. In Fig. \ref{fig:barchart_bzip/fernet_gzip/nacl}, we see that the CF \verb|bzip-then-Fernet| sees an almost 45-50\% increase in compression time when compressing first. This result may imply that the encryption with \verb|Fernet| decreases entropy on the file, thus making the compression easier, however this is pure speculation as this result is unintuitive otherwise. For \verb|gzip-then-NaCl| we see an increase of roughly 30-40ms in files of size 1MB (about double the time), and in a 10MB file this doubling becomes an increase of around 300ms. This, along with a few other algorithm pairs, leads us to conclude that in relatively few or rare cases the operation order with the fastest time can be EF.

\subsection{File Size Impacts}

\begin{figure*}
        \centering
    \includegraphics[width=0.22\textwidth]{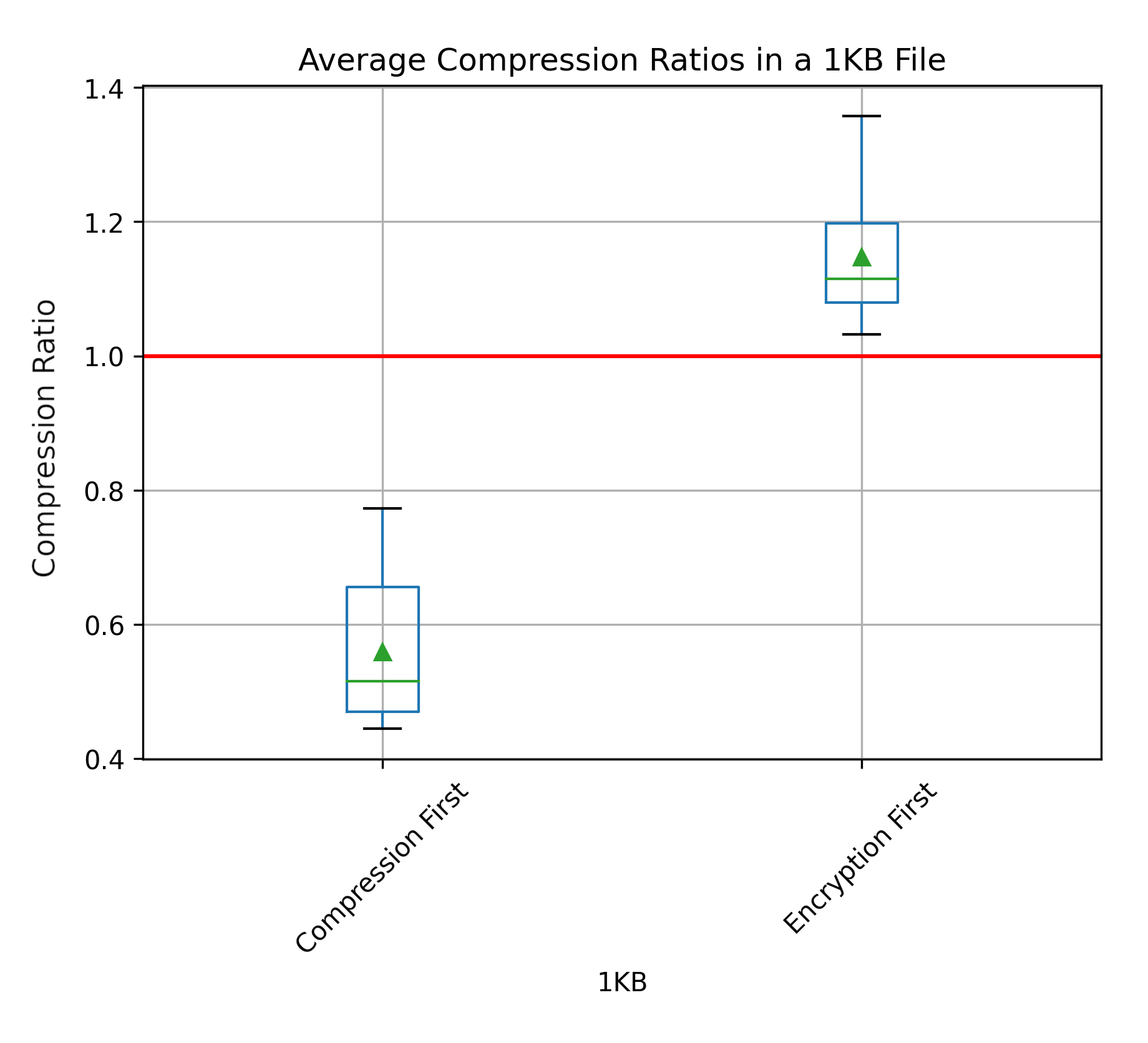}
    \includegraphics[width=0.22\textwidth]{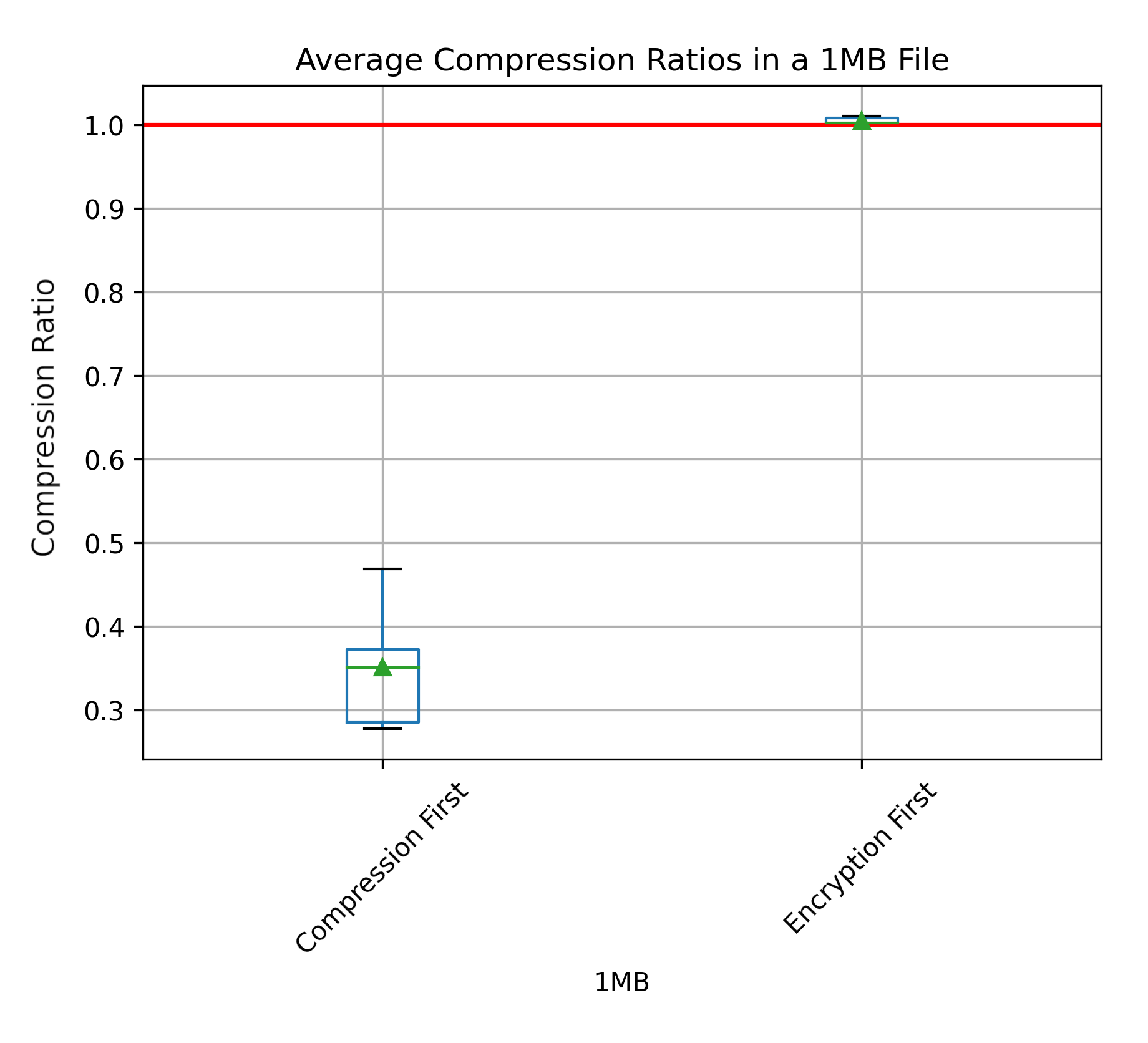}
    \includegraphics[width=0.22\textwidth]{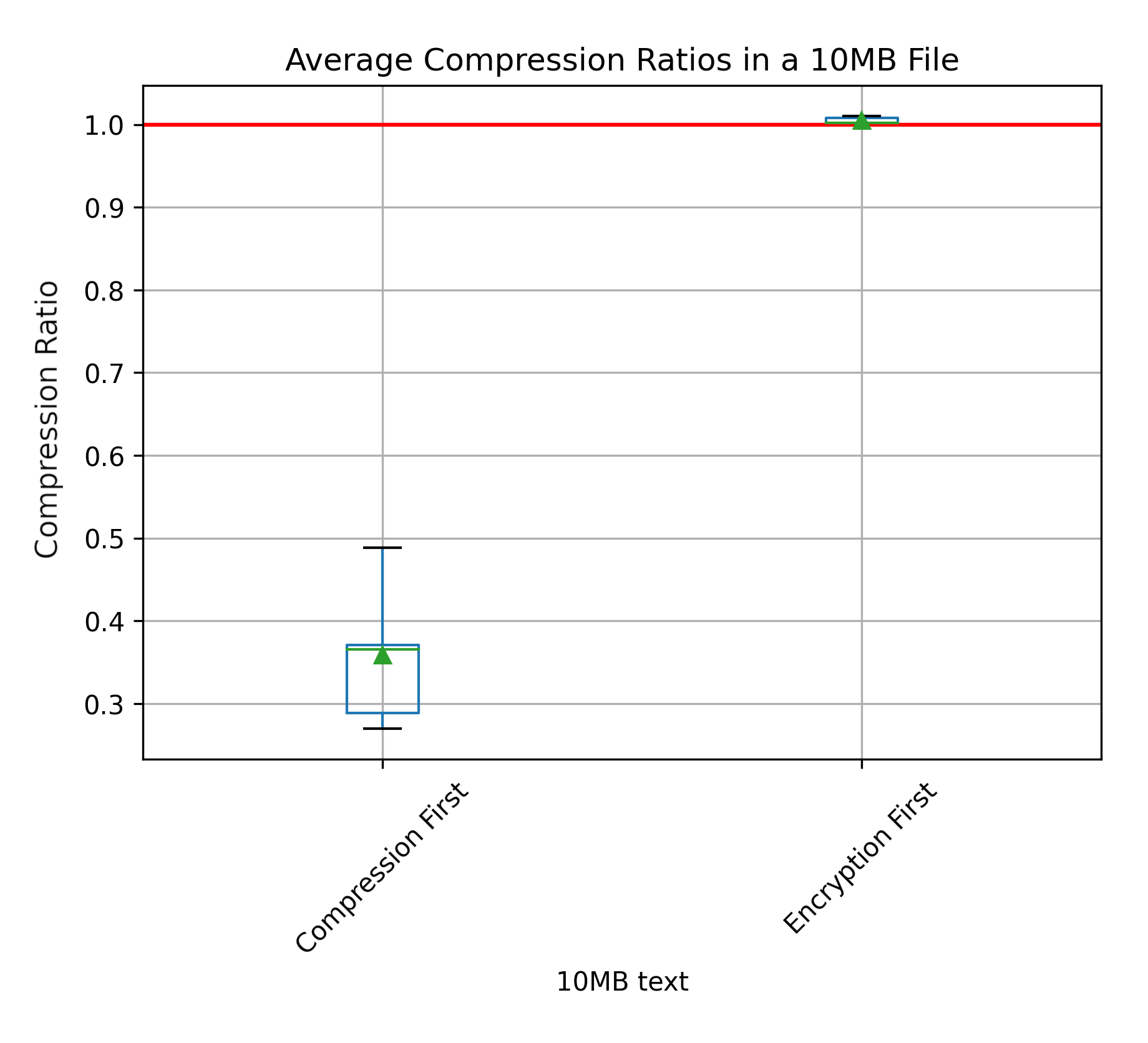}
    \includegraphics[width=0.22\textwidth]{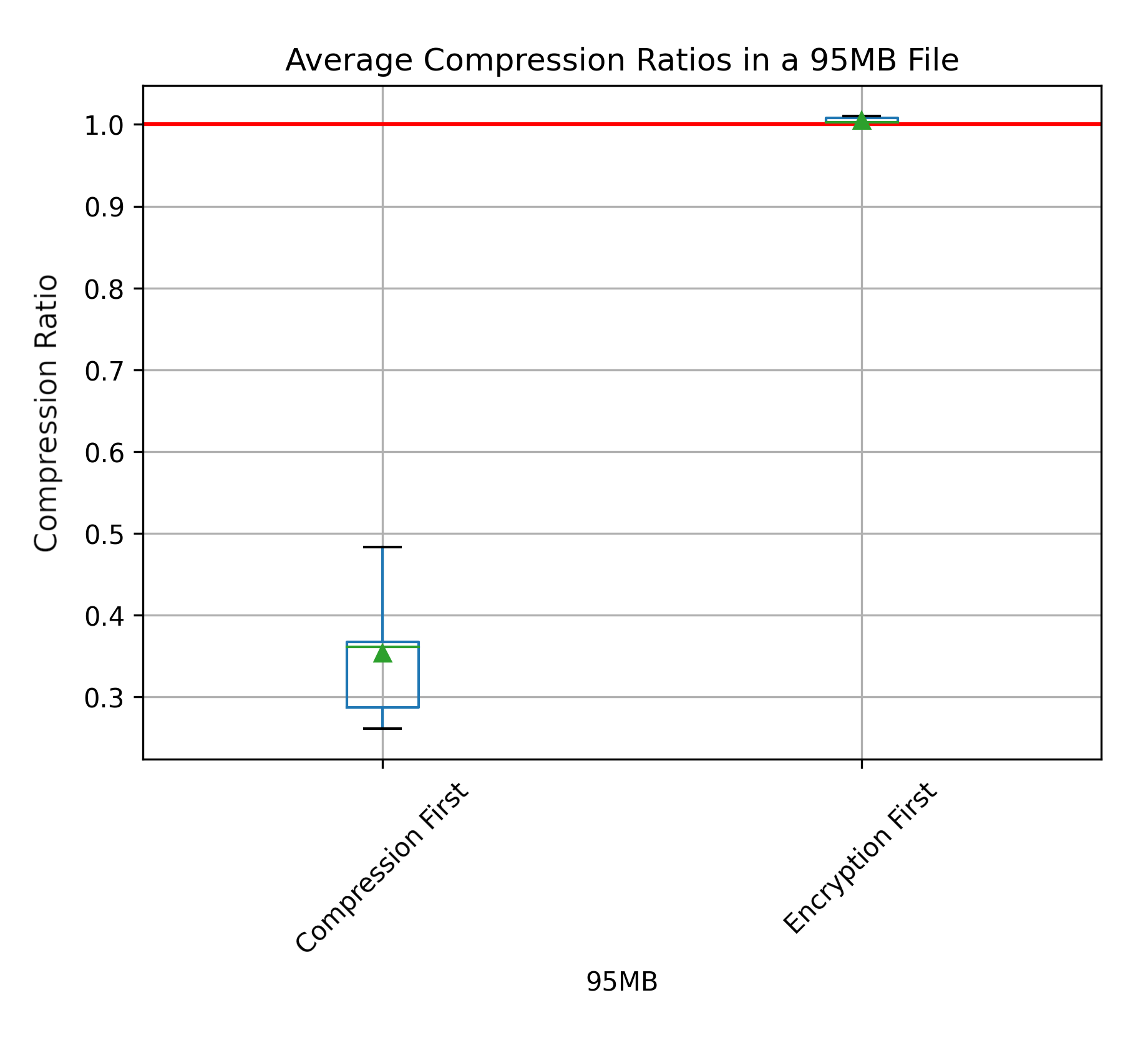}
    \caption{Compression ratios across several file sizes. We find that the smaller files have an overall worse compression ratio for both EF and CF approaches.}
    \label{fig:compress-ratios}
\end{figure*}

When considering the impact on data sizes, our results in Fig. \ref{fig:compress-ratios} also confirm CF's dominance. CF is able to achieve compression ratios that are below the original file size (indicated by the red line), and with significant reduction of 25-50\% at least when compared to EF. There is not context where EF produced final data size that was below the original size. This is intuitive as encryption would increase the size of data in most cases, and if done well compression may very well struggle as indicated in the related work section.

\section{Impact on 5G Teleoperations}
From our assessments of algorithm time and size performance, we now apply these observations to a practical use case: teleoperations over 5G.

For real-time transmission, the sum of delay from file read to file reception must be less than 100ms (0.1 seconds)\cite{timings}. For an example with a 10MB PDF file, we add a network latency variable between Operation 2 and Operation 2 Inverse (visualized in Figure \ref{fig:barchart_bzip/fernet_gzip/nacl}) to approximate network transmission time. When considering the fastest algorithm for file sizes $<$1MB, we observed files are encrypted, compressed, transmitted with a theoretical and static network latency of 50ms, decrypted, and decompressed in less than 100ms 100\% of time. In files with size 1MB, real-time transmission is demonstrated in 73\% of samples with the fastest algorithm pair. Lastly, in files $\geq$10MB, the fastest algorithm pair achieves real-time transmission 0\% of the time.

Broadly, this means that for larger size data transmissions, the algorithms may not be quite performant enough to apply encryption and compression together, requiring increases in capacity or intelligent decision systems.

\section{Limitations and Future Work}
This work can be considered a light but robust exploration of the interplay of compression and encryption. Further work can expand on the existing dimensions, including more algorithms and data types. Additionally, examining in detail the impact interplay has on the entropy and other security considerations may be examined as well.

\section{Conclusions}
In this paper, we conducted an assessment of the interplay of compression and encryption and its implications on 5G teleoperations. From these examinations, we confirmed that CF is generally the better performing operation order for compression and encryption, but that in certain algorithmic pairs the inverse may be true. Additionally, the impact of using both encryption and compression for files that are sufficiently large will make them unsuitable for real-time transmission, dampening the possibilities for 5G-teleoperations using secure and efficient transmission.

\bibliographystyle{ACM-Reference-Format}
\bibliography{bib.bib}

\end{document}